# CaR-FOREST: JOINT CLASSIFICATION-REGRESSION DECISION FORESTS FOR OVERLAPPING AUDIO EVENT DETECTION

*Huy Phan*[⋆†], *Lars Hertel*[⋆], *Marco Maass*[⋆], *Philipp Koch*[⋆], and *Alfred Mertins*[⋆]

[⋆]Institute for Signal Processing, University of Lübeck, Germany
[†]Graduate School for Computing in Medicine and Life Sciences, University of Lübeck, Germany
Email: {phan, hertel, maass, koch, mertins}@isip.uni-luebeck.de

## ABSTRACT

This report describes our submissions to Task2 and Task3 of the DCASE 2016 challenge [1]. The systems aim at dealing with the detection of overlapping audio events in continuous streams, where the detectors are based on random decision forests. The proposed forests are jointly trained for classification and regression simultaneously. Initially, the training is classification-oriented to encourage the trees to select discriminative features from overlapping mixtures to separate positive audio segments from the negative ones. The regression phase is then carried out to let the positive audio segments vote for the event onsets and offsets, and therefore model the temporal structure of audio events. One random decision forest is specifically trained for each event category of interest. Experimental results on the development data show that our systems significantly outperform the baseline on the Task2 evaluation while they are inferior to the baseline in the Task3 evaluation.

## 1. INTRODUCTION

Audio event detection is an important problem of computational auditory scene analysis [2, 3]. It addresses the problem of determining where in time an event is happening in an audio stream and the identity of that event. Therefore, it enables many novel applications such as ambient assisted living [4], security surveillance [5] to name a few.

Besides two common detection approaches, i.e., detection-by-classification [6, 7, 8] and automatic speech recognition (ASR) based frameworks [9, 10], a regression-based approach using random decision forests has been proposed in our previous works [11, 12, 13, 14]. Unlike the common approaches, the regression-based one accomplishes the detection task by letting the audio segments vote for the positions of event onsets and offsets. As a results, this inherently models the temporal structure of audio events [15]. This approach has been reported significantly more favorable performance compared to those of the common ones [12]. More interestingly, due to the monotonicity of its detection function [13], early event detection without losing detection accuracy can be guaranteed [13] and multi-channel fusion becomes straightforward [14].

Regarding detection of overlapping events, it has been shown in the literature that handling event overlap, e.g., multiple events that happen simultaneously in time, is very challenging. In intuition, when two events overlaps each other, their frequency contents are mixed up. Three different strategies have been proposed to deal with this. The first one aims at entangling the mixtures into the individual constituents using blind source separation [16, 17], the second is based on direct classification [8, 18, 19], and the third relies on the selection of local spectral information. While the regression-forest detectors are superior for nonoverlapping event detection, they are not supposed to handle well event overlaps. It is due to the fact that they are trained on only positive examples and know nothing about the negative ones [11, 12]. In order to overcome this, we additionally equip them with a feature selection capability. We propose discriminative decision forest detectors which are trained using both positive and negative examples. The forests are trained to deal with both classification and regression tasks at the same time. The classification-oriented training enforces the forests to select the discriminative features from polyphonic mixtures to separate positive examples from negative ones. The regression-oriented training then follows up to model temporal structures of audio events as in regular regression forests [11, 12].

## 2. DISCRIMINATIVE DECISION FOREST REGRESSION

We describe in this section the framework of discriminative decision forests that are jointly trained for the classification and regression tasks. This framework will be employed in our submissions for Task2 and Task3 of the DCASE 2016 challenge.

### 2.1. Joint classification-regression training algorithm

Given segmented audio events obtained from annotated training data, we label the event instances of the target category as positive examples and others (i.e. other event categories and background) as negative ones. The training examples are decomposed into the set of interleaved audio segments $\mathcal{S} : \{s_n = [\mathbf{x}_n, c, \mathbf{d}_n]; n = 1\ldots|\mathcal{S}|\}$ where $\mathbf{x}_n \in \mathbb{R}^M$ is the feature vector for the segment $n$, and $M$ is the dimensionality. $c \in \{0, 1\}$ denotes the class label inherited from the event instance that the segment stems from. $\mathbf{d}_n = [d_n^+, d_n^-] \in \mathbb{R}_+^2$ is the distance vector where $d_n^+$ and $d_n^-$ denote the distances from $n$ to the first segment (i.e. the onset) and the last segment (i.e. the offset) inclusive of the corresponding event respectively [11, 12]. Especially, for the negative examples, their distance vectors are not used during training and, therefore, can be ignored.

Suppose that we learn a forest model of $T$ trees. To construct a single decision tree, we randomly sample and use a subset from $\mathcal{S}$. Different from the regression-only forests in [11, 12, 15, 14], a split node of the tree performs either classification (therefore the name "discriminative") or regression tasks, for example by randomly deciding which task will be performed. The splitting function is defined as

$$t_{r,q,\tau}(\mathbf{x}) = \begin{cases} 1, & \text{if } x_r - x_q > \tau \\ 0, & \text{otherwise.} \end{cases} \quad (1)$$

Here, $x_r$ and $x_q$ denote the values of $\mathbf{x}$ at a random selected feature channels $r, q \in \{1, \ldots, M\}$, respectively. $\tau$ is a random threshold generated in the range of $x_r - x_q$. Opposed to the single feature test function in [11, 12], the test function in (1) is expected to better explore the internal structure of the feature space.

We further denote $S$ to be the set of audio segments that arrived at the current split node. Evaluating a test $t$ on $S$ will split it into two subsets $S_R$ and $S_L$ where $S_R = \{\mathbf{x} \in S | t(\mathbf{x}) = 1\}$ and $S_L = \{\mathbf{x} \in S | t(\mathbf{x}) = 0\}$. For the classification task, the quality of a test is defined by the information gain $I(\cdot)$ given by

$$I(t) = H(S) - \left(\frac{|S_R|}{|S|} H(S_R) + \frac{|S_L|}{|S|} H(S_L)\right) \quad (2)$$

where $|\cdot|$ denotes the size of a set and $H(\cdot)$ measures the purity of a set in terms of class labels. We use the entropy for $H(\cdot)$:

$$H(S) = -\sum_{c \in \{0,1\}} P(c) \log P(c). \quad (3)$$

For regression, the quality of a test is measured by the total distance variation:

$$V(t) = V_R + V_L, \quad (4)$$

where

$$V_R = \sum_i \|\mathbf{d}_i - \bar{\mathbf{d}}\|_2^2 \text{ for } \mathbf{x}_i \in S_R, \quad (5)$$

$$V_L = \sum_i \|\mathbf{d}_i - \bar{\mathbf{d}}\|_2^2 \text{ for } \mathbf{x}_i \in S_L. \quad (6)$$

In (5) and (6), $\bar{\mathbf{d}}$ denotes the mean distance vector of the set. Note that only positive examples are involved when training for regression.

We randomly generate a large number of tests from which an optimal one is then adopted to maximize $I(\cdot)$ or minimize $V(\cdot)$ depending on whether the classification or regression training is being performed. Eventually, $S_R$ and $S_L$ are further directed to the right and left child node, respectively. The process is repeated recursively until either a maximum depth $D_{\max}$ of the tree is reached or a minimum number $N_{\min}$ of audio segments are left. A leaf node is then created. The class label distribution of the audio segments that arrived at the leaf is then modeled as

$$P(\text{positive}) = \frac{|\{\mathbf{x} \in S | c = 1\}|}{|S|}, \quad (7)$$

$$P(\text{negative}) = \frac{|\{\mathbf{x} \in S | c = 0\}|}{|S|}. \quad (8)$$

Furthermore, the onset and offset distances of the positive audio segments that arrived at the leaf are modeled as Gaussian distributions:

$$\mathcal{N}^+(d; \bar{d}^+, \Sigma^+) = \frac{1}{\sqrt{2\pi\Sigma^+}} \exp\left(-\frac{(d - \bar{d}^+)^2}{2\Sigma^+}\right), \quad (9)$$

$$\mathcal{N}^-(d; \bar{d}^-, \Sigma^-) = \frac{1}{\sqrt{2\pi\Sigma^-}} \exp\left(-\frac{(d - \bar{d}^-)^2}{2\Sigma^-}\right). \quad (10)$$

Here, $(\bar{d}^+, \Sigma^+)$ and $(\bar{d}^-, \Sigma^-)$, respectively, denote the means and variances of onset and offset distances of the positive audio segments. Eventually, the saved models $P(\text{positive})$, $P(\text{negative})$, $\mathcal{N}^+(d; \bar{d}^+, \Sigma^+)$, and $\mathcal{N}^-(d; \bar{d}^-, \Sigma^-)$ are calibrated using the whole training data set.

The algorithm is repeated to construct all trees of the forest. Although a tree can be trained by alternatively performing classification and regression, we employ a steering parameter $\gamma$. At the level of a tree less than or equal $\gamma$, the training is classification-oriented. On contrary, at the level of a tree greater than $\gamma$, the training is targeted for regression.

### 2.2. Testing

Given a test audio signal, we want to estimate the onset and offset time of a target event. We input a segment $\mathbf{x}_m$ at the index $m$ into one tree of the trained forest. At every split node, the stored binary test is evaluated, directing it to the right or left child until ending up at a leaf node. The models $P(\text{positive})$, $P(\text{negative})$, $\mathcal{N}^+(d; \bar{d}^+, \Sigma^+)$, and $\mathcal{N}^-(d; \bar{d}^-, \Sigma^-)$ stored at the leaf are retrieved. We then obtain estimates for the onset and offset positions at a time index $n$ as

$$p^+(n|\mathbf{x}_m, \bar{d}^+, \Sigma^+) = P(\text{positive}|\mathbf{x}_m)\mathcal{N}^+(n; m - \bar{d}^+, \Sigma^+), \quad (11)$$

$$p^-(n|\mathbf{x}_m, \bar{d}^-, \Sigma^-) = P(\text{positive}|\mathbf{x}_m)\mathcal{N}^-(n; m + \bar{d}^-, \Sigma^-). \quad (12)$$

Moreover, we only allow an audio segment $\mathbf{x}_m$ with a classification probability $P(\text{positive}|\mathbf{x}_m) \geq \alpha$ to contribute on the estimation in (11) and (12). The determination of the threshold $\alpha$ will be explained later. The estimation by the forest is computed by averaging over all trees $i$:

$$p^+(n|\mathbf{x}_m) = \frac{1}{T} \sum_{i=1}^{T} p^+(n|\mathbf{x}_m, \bar{d}_i^+, \Sigma_i^+), \quad (13)$$

$$p^-(n|\mathbf{x}_m) = \frac{1}{T} \sum_{i=1}^{T} p^-(n|\mathbf{x}_m, \bar{d}_i^-, \Sigma_i^-). \quad (14)$$

The estimates by all segments are accumulated to yield the confidence scores for the onset and offset positions of the target event as

$$f^+(n) = \sum_m p^+(n|\mathbf{x}_m), \quad (15)$$

$$f^-(n) = \sum_m p^-(n|\mathbf{x}_m). \quad (16)$$

A threshold $\beta$ is finally applied on the confidence scores $f^+(n)$ and $f^-(n)$ as in [12, 14] to identify the event onsets and offsets. As long as we find a pair of maximum confidence scores above the detection threshold $\beta$ in chronological order, a maximum onset score followed by the maximum offset score, a target event is considered detected.

## 3. TASK 2: EVENT DETECTION IN SYNTHETIC DATA

This task focuses on event detection of overlapping office sounds in synthetic mixtures. The training data is given by 20 isolated event instances for each of eleven target categories. The development data is created by adding the given isolated events into continuous recordings of background noise at various signal-to-noise (SNR) levels (i.e. -6, 0, and 6 dB), event density conditions, and polyphony. The test data is created with a similar manner but with unseen isolated events.

Table 1: Task2 overall performance on the development data and test data.

| | Development set | | | | Test set | | | |
|---|---|---|---|---|---|---|---|---|
| | Segment-based | | Event-based | | Segment-based | | Event-based | |
| | ER | F1 | ER | F1 | ER | F1 | ER | F1 |
| **Our system** | 0.1420 | 92.8 | 0.1835 | 90.5 | 0.5901 | 64.8 | 1.0123 | 39.8 |
| **Baseline** | 0.7859 | 41.6 | - | 30.3 | 0.8933 | 37.0 | 1.6852 | 24.2 |

## 3.1. Experiments on the development data

### 3.1.1. Training data creation

Firstly, isolated training examples were scaled relatively to the background noise in the development data to result in three replica corresponding to -6, 0, and 6 dB. We also created artificial overlapping events from the isolated ones. Each positive event instance was mixed with randomly selected negative event instances, one from each negative event category, at a random position with an overlap of at least 50% of its duration. These overlapping instances were labeled as positive examples. In addition, we created negative overlapping instances by mixing a negative event instance with another randomly selected negative one. The artificial mixing procedure was performed for each SNR level.

### 3.1.2. Low-level features

The training and development data were downsampled to 16 kHz. The signals were then decomposed into 100 ms segments with an overlap of 90 ms. We extracted 64 Gammatone ceptral coefficients [20] in the frequency range of 50-8000 Hz to represent each segment. For the development data, we conducted background noise subtraction [21] before feature extraction. Furthermore, since the training data does not contain background noise, we randomly selected a number of background segments (equal to the number of positive audio segment) from the development data to add them into the training set.

### 3.1.3. Parameters

We trained a discriminative decision forest for each event category using the algorithm presented in Section 2. The number of trees was set to ten. A random subset containing 50% training audio segments was used for training. During training, 20,000 binary tests were generated for a split node. In addition, we set the maximum depth to $D_{\max} = 12$ and minimum number of segments at leaf nodes to $N_{\min} = 20$. Furthermore, we opted the steering parameter $\gamma = 9$.

The confidence scores outputed by the forests were normalized by dividing by the maximum scores in development data. The positive classification probability threshold $\alpha$ and the detection threshold $\beta$ were found by a grid search to minimize cross-validation segment-based total error rate (ER). The threshold $\alpha$ was searched in the range of [0, 1] with a step size of 0.05 and $\beta$ was searched in the range of [0, 1] with a step size of 0.025.

### 3.1.4. Experimental results

The detection performance of our system is shown in Table 1. Overall, a segment-based ER (the main evaluation metric for the task) of 0.1420 is obtained by our system on the development which outperforms the baseline by a large margin of 0.6439 absolute. Furthermore, significant improvements can be seen on other metrics.

Table 2: Task2 segment-based class-wise performance on test data.

| Event Type | Our system | | Baseline | |
|---|---|---|---|---|
| | ER | F1 | ER | F1 |
| Clearing throat | 0.6506 | 62.7 | 0.8956 | 49.4 |
| Coughing | 0.8491 | 47.2 | 1.0561 | 6.2 |
| Door knock | 0.6595 | 73.1 | 0.8674 | 63.1 |
| Door slam | 1.3092 | 15.6 | 2.9855 | 15.1 |
| Drawer | 0.7433 | 62.1 | 0.9833 | 3.3 |
| Keyboard | 0.4599 | 79.5 | 0.9100 | 22.4 |
| Keys (put on table) | 0.7956 | 40.5 | 0.6400 | 62.7 |
| Human laughter | 0.7199 | 64.1 | 1.3193 | 40.8 |
| Page turning | 0.5929 | 62.9 | 1.0032 | 2.5 |
| Phone ringing | 0.4136 | 77.8 | 0.7956 | 40.6 |
| Speech | 0.5626 | 71.7 | 0.7163 | 59.3 |
| **Average** | 0.7051 | 59.7 | 1.1066 | 33.2 |

## 3.2. The final submission system and results

The detection system was finally evaluated on the test data to produce the final results for submission. However, since the development data is made of training isolated events, there exists a significant mismatch between the confidence scores obtained by the development data and the test data. Specifically, the confidence scores obtained by the development data are much higher than those of the test data most likely due to overfitting. We mitigated this effect by average filtering the scores with a window size of eleven audio segments (equivalent to 100 ms). Finally, we rejected detected events that have durations longer than thrice of the maximum event duration in the training data.

As shown in Tables 1 and 2, our system achieves an overall and class-wise segment-based ERs of 0.5901 and 0.7051, respectively, on the test data. These results outruns those of the baseline by 0.3032 and 0.4015 absolute.

The significant performance drop on the test data compared to that on the development data is likely due to overfitting given the fact that the development recordings are composed of the training examples.

## 4. TASK 3: EVENT DETECTION IN REAL LIFE DATA

This task focuses on audio event detection in real life data [1, 22]. The data consists of recordings from two acoustic scenes: home

Table 3: Task3 overall performance on development and test data.

| | Dev. set | | Test set | | | |
| --- | --- | --- | --- | --- | --- | --- |
| | Segment-based | | Segment-based | | Event-based | |
| | ER | F1 | ER | F1 | ER | F1 |
| **Our system** | 0.8304 | 31.6 | 0.9644 | 23.9 | 1.0634 | 1.5 |
| **Baseline** | 0.9100 | 23.7 | 0.8773 | 34.3 | 1.7303 | 6.3 |

(indoor) and residential area (outdoor). Eleven and seven event categories are targeted for the home and residential area datasets, respectively. Each dataset is partitioned into four folds for development. The overall detection over all folds will be reported. Our experiments were repeated for each fold and each acoustic scene.

### 4.1. Training data and low-level features

We extracted segmented positive audio event instances from the annotated data. The rest outside the duration of positive instances were considered as negative examples. The signals were first downsampled to 16 kHz and then decomposed into 50 ms segments with an overlap of 40 ms. We extracted 64 Gammatone ceptral coefficients [20] in the frequency range of 20-8000 Hz to represent a segment.

### 4.2. Parameters

The parameters used to train the decision forests are the same as those in the Task 2 system, except for the following difference. We employed a thresholding scheme called *ignorance threshold*. Based on an observation for some event classes that the error rates $ER > 1.0$. That is, for these classes, doing nothing is better something. This fact can be easily introduced into our system to guarantee the maximum error rate of 1.0 by simply allowing the detection threshold $\beta$ to be larger than 1.0 during the parameter search. If $\beta > 1.0$ is found, the detection system will be unable to take effective action since the threshold always stays higher than the confidence scores.

On another hand, for the home scene, the confidence scores are smoothed by average filtering with a window size of eleven audio segments (equivalent to 100 ms) in order to reduce outliers.

### 4.3. Experimental results on the development data

The overall performance of our detection systems on the development data are presented in Tables 3. As can be seen, our systems obtain an overall segment-based ER and F1-score of 0.8304 and 31.6% (averaged over *Home* and *Residential area data*), which outperform the baseline by about 0.0796 and 7.9% absolute, respectively.

### 4.4. The final submission system and results

The final detection system was trained with the same settings as that for the development data but we used the whole development data for training. The trained system was then evaluated on the test data to produce the final results.

As can be seen from Table 3, our system is inferior to the baseline (i.e. 0.9644 compared to 0.8773 on the overall segment-based ER). On class-wise performance, although our system obtains an absolute gain of 0.2612 in terms of average segment-based ER over the baseline on the *Residential area* data (as shown in Table 5), it is subordinate to the baseline with an absolute loss of 0.0666 on the *Home* data (as shown in Table 4).

Table 4: Task3 (Home) segment-based class-wise performance on test data.

| Event Type | Our system | | Baseline | |
| --- | --- | --- | --- | --- |
| | ER | F1 | ER | F1 |
| Cupboard | 1.0769 | 6.7 | 1.0385 | 0.0 |
| Cutlery | 1.3143 | 0.0 | 1.0571 | 0.0 |
| Dishes | 1.0000 | 0.0 | 1.0744 | 15.6 |
| Drawer | 1.0000 | 0.0 | 0.9811 | 7.1 |
| Glass jingling | 1.1333 | 0.0 | 1.0000 | 0.0 |
| Object impact | 1.0000 | 2.0 | 1.1574 | 12.3 |
| Object rustling | 1.0000 | 0.0 | 0.6786 | 59.6 |
| Object snapping | 1.0000 | 0.0 | 1.0000 | 0.0 |
| People walking | 1.0000 | 0.0 | 1.0833 | 16.1 |
| Washing dishes | 1.0000 | 0.0 | 1.0190 | 0.0 |
| Water tap running | 0.9693 | 31.1 | 0.6724 | 64.0 |
| **Average** | 1.0449 | 3.6 | 0.9783 | 15.9 |

Table 5: Task3 (Residential area) segment-based class-wise performance on test data.

| Event Type | Our system | | Baseline | |
| --- | --- | --- | --- | --- |
| | ER | F1 | ER | F1 |
| Bird singing | 1.4673 | 43.7 | 0.9637 | 30.9 |
| Car passing by | 0.8263 | 46.3 | 0.4836 | 74.8 |
| Children shouting | 1.0000 | 0.0 | 1.1333 | 0.0 |
| Object banging | 1.0000 | 0.0 | 1.0000 | 0.0 |
| People speaking | 1.0000 | 0.0 | 2.6667 | 1.3 |
| People walking | 1.1096 | 0.0 | 1.1096 | 6.9 |
| Wind blowing | 1.0000 | 0.0 | 1.8750 | 25.0 |
| **Average** | 1.0576 | 12.9 | 1.3188 | 19.8 |

## 5. CONCLUSIONS

We present in this report our audio event detection systems for Task2 and Task3 of the DCASE 2016 challenge. We propose to train discriminative decision forests that are jointly learning for the classification and regression tasks. With the classification-oriented training, the forests are able to select the discriminative features in the mixtures of event overlaps to separate positive examples from negative ones. Furthermore, the forests can also model the temporal structure of audio events due to regression-oriented training. We learn such a decision forest for each of target event category for the detection purpose. Experimental results on the development data show that our systems outperform the DCASE 2016 challenge baselines on Task2 evaluation but is inferior to the baseline on Task3 evaluation.

## 6. ACKNOWLEDGEMENT

This work was supported by the Graduate School for Computing in Medicine and Life Sciences funded by Germany's Excellence Initiative [DFG GSC 235/1].